\documentclass{article}
\usepackage{amssymb}
\usepackage{amsmath}
\usepackage[numbers]{natbib}
\usepackage{rotating}
\usepackage{amsmath}
\usepackage{bbm}
\usepackage{float}
\usepackage{graphicx}
\usepackage{enumitem}
\usepackage[margin=1in]{geometry}
\begin{document}

\title{Score-Preserving Targeted Maximum Likelihood Estimation}
\author{\makebox[.9\textwidth]{Noel Pimentel}\\Division of Biostatistics\\ UC Berkeley \and Alejandro Schuler\\Division of Biostatistics\\ UC Berkeley \and Mark van der Laan\\Division of Biostatistics\\ UC Berkeley }

\maketitle

\begin{abstract}
Targeted maximum likelihood estimators (TMLEs) are asymptotically optimal among regular, asymptotically linear estimators. In small samples, however, we may be far from ``asymptopia'' and not reap the benefits of optimality. Here we propose a variant (score-preserving TMLE; SP-TMLE) that leverages an initial estimator defined as the solution of a large number of possibly data-dependent score equations. Instead of targeting only the efficient influence function in the TMLE update to knock out the plug-in bias, we also target the already-solved scores. Solving additional scores reduces the remainder term in the von-Mises expansion of our estimator because these scores may come close to spanning higher-order influence functions. The result is an estimator with better finite-sample performance. We demonstrate our approach in simulation studies leveraging the (relaxed) highly adaptive lasso (HAL) as our initial estimator. These simulations show that in small samples SP-TMLE has reduced bias relative to plug-in HAL and reduced variance relative to vanilla TMLE, blending the advantages of the two approaches. We also observe improved estimation of standard errors in small samples.
\end{abstract}

\section{Notation and Preliminaries}

We begin by reviewing the basics of our estimation problem, efficiency theory, and targeted maximum likelihood. More detailed introductions can be found in \cite{imci, tmle2018, kennedy2023semiparametricdoublyrobusttargeted, fisher2019visuallycommunicatingteachingintuition, Hines_2022}.

\subsection{Setup}

Presume we observe $n$ observations $O_1,\dots,O_n \overset{\mathrm{i.i.d.}}{\sim} P_0$ IID from the same distribution $P_0$ contained in a set of distributions $\mathcal M$ encoding our assumptions. Let $P_n$ denote the empirical probability measure that puts mass $1/n$ at each $O_i$. In our running example we will take $O = (W,A,Y)$, a $d$-dimensional vector of covariates $W$, a given binary treatment $A$, and an observed binary outcome $Y$.

We are interested in a ``pathwise-differentiable'' \cite{luedtke2023onestepestimationdifferentiablehilbertvalued, UHAL} target parameter $\Psi: \mathcal{M} \mapsto \mathbbm{R}^d$. For example, we may take the treatment-specific mean $\Psi(P_0) = E_{W,0}[E_0[Y|A=1,W]]$ which under standard assumptions identifies the average population outcome under treatment \cite{imci, tmle2011, tmle2018}. 

At each point $P$ of $\mathcal M$ (say with density function $p$ w.r.t. some dominating measure) we define a set of \textit{scores}. Informally, each score at $P$ is the initial rate of change in the log-likelihood ($h(P) = \frac{d}{d\epsilon} \log p_\epsilon \big|_{\epsilon=0}$) as we progress along a given one-dimensional submodel (path), e.g. $p_\epsilon = \epsilon p + (1-\epsilon)q$ where $q$ is some other density in the model. This assumes the model is convex so that such paths are in the model. A score $h(P)$ might be thought of as an (generally infinite-dimensional) ``direction'' (in the sense of North, East, etc.) in which we can move away from the density $p$ and remain in $\mathcal M$.

\subsection{Efficiency Theory}
An estimator $\hat\Psi(P_n)$ is a mapping from observed data to a real number. An estimator is called \textit{asymptotically linear} at P with influence function $D(P)$ if $\hat\Psi(P_n) - \Psi(P) = \frac{1}{n} \sum D(P)(O_i) + o_P(n^{-1/2})$. The notation $D(P)$ indicates that the influence function of an estimator depends on the distribution of the data $P$. 
Plainly, asymptotic linearity means that the estimator is consistent and behaves like an average of IID random variables $D(P)(O_i)$ in large samples. By the central limit theorem, the asymptotic distribution of an asymptotically linear estimator with influence function $D(P)$ is $\sqrt{n}(\hat\Psi(P_n) - \Psi(P)) \rightsquigarrow N(0, V[D(P)(O)])$, which facilitates the construction of confidence intervals, p-values, etc.

A central result of efficiency theory establishes that among regular, asymptotically linear (RAL) estimators for a pathwise-differentiable parameter, there is a minimum-variance influence function $D^*(P)$ which we call the \textit{efficient} influence function (EIF). An estimator is called \textit{efficient} if it has the efficient influence function. An efficient estimator is asymptotically optimal in the sense that no other RAL estimator can be expected to give lower estimation error for large samples. The same result establishes that the EIF is a \textit{score} in our model at $P$ (indeed it is the only influence function that is also a score). Moreover moving locally within $\mathcal M$ away from $P$ in the direction $D^*(P)$ produces the largest rate of change in $\Psi(P)$ among possible directions.

The efficient influence function can be derived analytically for many parameters of interest. For example, for the treatment-specific mean the efficient influence function is $\frac{A}{P(A=1|W)}(Y-E_P[Y|A=1,W]) + (E_P[Y|A=1,W] - E_P[E_P[Y|A=1,W]])$. The subscripts explicitly show the dependence on $P$. 

\subsection{Targeted Maximum Likelihood Estimation}
We say an estimator is a \textit{plug-in} if it can be written as $\hat\Psi(P_n) = \Psi(\hat P_n)$ where $\hat P_n \in \mathcal M$ is an estimate of the distribution derived from the data. If we define $R = \Psi(\hat P_n) - \Psi(P_0) + PD^*(\hat P_n) $, for any plug-in estimator, we can expand it by definition \cite{imci, tmle2011, tmle2018}\footnote{Above and throughout the paper we adopt the empirical process notation $P f = \int f(Z) \, dP$ and $P_n f = \frac{1}{n}\sum_i f(Z_i)$. In this notation these operators do not average over any potential randomness in $f$ so $P f_n$ is a random variable if $f_n$ is a random function learned from data.} and get the following decomposition
\begin{equation}
\label{eq:expansion}
\Psi(\hat P_n) - \Psi(P_0)
= 
P_n D^*(P_0) 
- P_n D^*(\hat P_n)
+ (P_n-P_0)(D^*(\hat P_n) - D^*(P_0) )
+ R
\end{equation}
 The first term in this expansion is the desirable one: if the other terms were $o_P(n^{-1/2})$, we would have that our plug-in is asymptotically linear with the efficient influence function. The last two terms can typically be shown to be $o_P(n^{-1/2})$ by design (using cross-fitting) and under generic assumptions (rate conditions on the convergence of nuisance function estimates) \cite{imci, tmle2011, tmle2018, Vaart_efficiency}, so the problem term is the ``plug-in bias'' $- P_n D^*(\hat P_n)
$. 

Targeted maximum likelihood estimation (TMLE) is a recipe for constructing an efficient plug-in estimator of a given pathwise differentiable parameter. To construct the TMLE we begin with an initial estimate of our distribution which we call $\hat P_n$. we then update that estimate by maximizing the empirical likelihood along the ``hardest'' submodel: any one-dimensional submodel $\{\tilde P_\epsilon: \epsilon \in \mathbb R\} \subset \mathcal M$ including $\hat P_n = \tilde P_{\epsilon=0}$ such that the score equation for this model is $D^*(\hat P_n)$. This is a one-dimensional optimization problem. We then repeat until convergence at an estimate $\hat P_n^*$. The final estimate of the target parameter is the plug-in $\Psi(\hat P_n^*)$.

The purpose of the updates is to eliminate the plug-in bias. If a point $P$ maximizes the likelihood in a given direction $h$, then the initial rate of change of the empirical log-likelihood along this path (i.e. the empirical mean of the score) is zero because we are at a local maximum in this direction at this point. Since we always move along paths with the score $D^*(\hat P_n)$ (updating $\hat P_n$ as we go), we have at convergence that $P_n D^*(\hat P_n^*) = 0$ and we have exactly eliminated the plug-in bias. 

Generically, we say that an estimate $\hat P_n$ \textit{solves} a score equation $h(\cdot)$ if $P_n h(\hat P_n) = 0$. An estimate $\hat P_n$ that solves a score equation $h$ is at a local maximum of the empirical log-likelihood in the ``direction'' $h(\hat P_n)$. TMLE eliminates the plug-in bias by solving the efficient influence function (one specific score): the estimate is at a point of local maximum likelihood in the \textit{targeted} direction, but not necessarily in other directions.

In the case of the treatment-specific mean for a binary outcome, the only part of the initial distributional estimate $\hat P_n$ that is updated by targeting is the initial estimate of the conditional expectation $\hat Q_n(1,W) = E_{\hat P_n}[Y|A=1,W]$ \cite{imci, tmle2011, tmle2018}. Moreover there is a particular submodel with score equal to the efficient influence function such that the TMLE converges in one step \cite{imci, tmle2011,tmle2018}. The update in this submodel is given by
\begin{equation}
\label{eq:update}
\text{logit}\, \hat Q_n^* = 
\text{logit}\, \hat Q_n + 
\hat\epsilon
\frac{A}{\hat g_n(W)}
\end{equation}
where $\hat g_n(W) = E_{\hat P_n}[A=1|W]$ is an estimated propensity score and $\hat\epsilon$ maximizes the empirical log-likelihood. Given the display, the optimal value of $\hat\epsilon$ is obtained by doing a logistic regression of $Y$ onto the ``clever covariate'' $\frac{A}{\hat g_n(W)}$ with a fixed offset $\text{logit}\, \hat Q_n$ and taking the resulting coefficient on the clever covariate. The final estimate is given by the plug-in $P_n \hat Q_n^*$. Therefore all we need to calculate the targeted estimate are the initial regression estimates $\hat Q_n$ and $\hat g_n$.

\subsubsection{Score-Solving Initial Estimators}

Many regression estimators are constructed to solve score equations. For example, consider fitting a linear regression 
\begin{equation}
\label{eq:regression}
    \hat Q_n(1,W) =
    \Phi_n(W)^\top \hat\beta
\end{equation}
by least squares where $\Phi_n: \mathbb R^d \to \mathbb R^{\tilde d}$ is some possibly data-dependent and high-dimensional feature expansion (e.g. a ploynomial regression or spline estimator). The minimizer $\hat \beta$ of the mean squared error satisfies $\frac{1}{n}\sum \Phi_n(W)(Y-\Phi_n(W)^\top\hat\beta) = 0$, i.e. it solves these $\tilde d$ score equations.

Solving scores can be very beneficial. If we believe in the regression model, solving the scores gives us a (general) maximum likelihood estimator in that model which can be used as an efficient plug-in for any smooth target parameter without the need for targeting \cite{Vaart_efficiency}. However, solving scores is also beneficial in nonparametric models. If the number and diversity of the solved scores increases with sample size, we are asymptotically better able to approximate the efficient influence function with the linear span (in $\mathcal L_2$) of the solved scores. It is immediate from the definition that any score that is in the span of solved scores is itself solved. Effectively, one can (nearly) solve the efficient influence function and eliminate plug-in bias (at a root-$n$ rate) without any explicit targeting. Indeed the scores solved by the initial estimator may begin to approximate any number of efficient influence functions, making the resulting distributional estimate simultaneously efficient as a plug-in for a large number of target parameters. This phenomenon underlies the ``undersmoothing'' approaches to efficient estimation \cite{UHAL, shi2024halbasedpluginestimationcausal, UndersmoothedEst, UHAL2}. 

Besides yielding asymptotic efficiency in some cases, solving a large number of scores can also improve finite-sample performance. Given the error expansion in equation \ref{eq:expansion}, the finite-sample performance of a plug-in generally depends on exactly how quickly the remainder term $R$ converges to zero. Previous works on higher-order efficient estimation diminish this term by constructing estimators that solve \textit{higher-order} efficient influence functions\cite{vanderlaan2021higher, Robins_2008}. Deriving these functions is difficult and entails computing pathwise derivatives of a sequential composition of functions. The resulting estimators are usually bespoke and brittle. However, the same goal can be accomplished by solving many scores, leveraging the same mechanism for establishing efficiency: as long as the space of solved scores is large enough, the higher-order efficient influence functions will be well-approximated by elements in this space and therefore solved to high precision. The result is an estimator with smaller error in small samples \cite{vanderlaan2021higher}.

In practice, constructing an efficient plug-in estimator using undersmoothing techniques is difficult: there is no single good way to do this. Therefore TMLE is still a prudent way to ensure efficiency. Unfortunately however, the benefits of score-solving initial estimators are not guaranteed to be preserved by the TMLE update step. Performing the update may solve the efficient influence function almost exactly, but at the cost of ``un-solving'' any scores already solved and harming small-sample performance. In what follows we present a simple modification of the TMLE algorithm that allows for the preservation of scores solved by an initial estimator. 

\section{Score-Preserving TMLE}

Our approach to construct a score-preserving TMLE (SP-TMLE) is generic and natural. The first step is to identify the score equations $h_j$ solved by the initial estimator $\hat P_n$. Once these are identified, we perform a \textit{multi-dimensional} TMLE update that targets these scores in addition to the usual efficient influence function and then iterate the update process.

TMLE can naturally handle multidimensional updates. Instead of maximizing the likelihood in a one-dimensional parametric submodel we use a $k$-dimensional submodel $\{\tilde P_\epsilon: \epsilon \in \mathbb R^k\} \subset \mathcal M$ again such that $\hat P_n = \tilde P_{\epsilon=0}$. Instead of a one-dimensional optimization we now have a $k$-dimensional optimization but this does not fundamentally change the process of empirically maximizing the likelihood in this model.\footnote{This idea can also be used to construct TMLEs that target multiple parameters simultaneously. The advantage of doing this in a plug-in framework is that, for example, estimates of mutually exclusive probabilities will sum to one.}

In the $k$-dimensional submodel we can ``move'' away from our initial estimate in the $k$ different directions corresponding to $\epsilon_1 \dots \epsilon_k$ \cite{ulfods, tmle2018_ch5}. These parameters map to $k$ scores $\frac{\partial}{\partial \epsilon_k} \log \tilde p_{\epsilon} \big|_{\epsilon=0}$. To knock out the plug-in bias at convergence we just need to ensure that the efficient influence function $D^*(\hat P_n)$ is spanned by these scores (so that it is solved at a point of empirical maximum likelihood). In our SP-TMLE we additionally ensure (by construction) that in each update step the scores $h_j(\hat P_n)$ are also spanned by our chosen submodel.

\subsection{SP-TMLE for the Treatment-Specific Mean}

As an example, consider estimating the treatment specific-mean, leveraging an initial estimator of the form $\hat Q_n(1,W) = \text{expit}(\Phi_n(W)^\top \hat\beta$). As long as the dimension $\tilde d$ of $\beta$ is less than $n$, we know that this estimator exactly solves the scores $h_j(\hat P_n) = \Phi_{n,j}(W) (Y - \Phi_n(W)\hat\beta)$ for $j \in 1\dots \tilde d$ where $\Phi_{n,j}$ is the $j$th basis function in the expansion $\Phi_n$. It is easy to show that the following $(\tilde d + 1)$-dimensional parametric submodel for $\hat Q_n$ spans each of these scores as well as the efficient influence function (when $\hat P_n(W)$ is given by the empirical distribution of the covariates):
\begin{equation}
\label{eq:sp-update}
\text{logit}\, \hat Q_n(\epsilon) = 
\text{logit}\, \hat Q_n 
+ \epsilon_{0} \frac{A}{\hat g_n(W)}
+ \sum_j^{\tilde d} \epsilon_j \Phi_{n,j}(W)  
.
\end{equation}
One way to obtain the coefficients for this update would be to perform a logistic regression of $Y$ onto the clever covariate and the basis functions $\Phi_{n,j}$ using an offset $\text{logit}\, \hat Q_n$. However, if $\tilde d \approx n$ the estimate may become unstable so instead we suggest incrementally iterating along the \textit{uniformly least-favorable submodel}. 

The idea here, which follows \cite{ulfods}, is to construct a submodel $\tilde P_\epsilon$ such that at each point $\epsilon$ in the submodel we have the desired scores $h_j(\tilde P_\epsilon)$ and $D^*(\tilde P_\epsilon)$. Generic submodels constructed to have score $h_j(\tilde P_{\epsilon=0})$ at $P_{\epsilon=0}$  typically do not have scores $h_j(\tilde P_{\epsilon})$ at generic points $P_{\epsilon}$. The purpose of having such a path is that we can theoretically perform the update in a single step. Additional exposition can be found in \cite{imci, tmle2011, tmle2018}.

In practice such models are difficult to construct explicitly. Nonetheless we can approximate the ``one-step'' update by taking many very \textit{small} steps, always following the direction of the score(s) we wish to solve. In our setting, we take a small value $\delta$ and let 
\begin{equation}
\label{eq:ulfm}
\epsilon = \delta (P_n S_n) \|P_n S_n\|^{-1}
\end{equation}
for the $(\tilde d + 1)$-dimensional vector $S_n(Y,A,W) = P_n\begin{bmatrix} 
A/\hat g_n(W) \\ \Phi_n(W) \end{bmatrix} 
\left( Y-\hat Q_{n}(W) \right)$. We plug the value of $\epsilon$ given in equation \ref{eq:ulfm} into equation \ref{eq:sp-update} to update $\hat Q_n$ and iterate until each of these  score equations are solved to a user-specified level of precision.

\section{Simulation Study}

Here we demonstrate in a simulation study that SP-TMLE improves TMLE's performance in small samples. The purpose of this study is to empirically validate our hypothesis in a simple, intelligible setting.

\subsection{Data-Generating Process}

Our simulation consists of data with structure $O = (W,A,Y)$. $W = (W_1, W_2)$ represents a vector of continuous baseline covariates, $A$ represents a vector of the exposure, and $Y$ represents a vector of the outcome. Each observation is independent and identically distributed. The following process generates the observed data:
\begin{align*} 
W_1 &\sim \text{Unif}(-1,1)\\
W_2 &\sim \text{Unif}(-1,1)\\
A &\sim \text{Bern}(g_0(A|W))\\
Y &\sim \text{Bern}({Q_0}(A,W))
\end{align*} 
the outcome mechanism is given by:
\begin{align*}
{Q}_0(A,W) &= \text{expit}(W_1 + 2W_2)
\end{align*} 
with value of the target parameter (treatment specific mean, not ATE): $$\Psi(P_0) = E_{W,0}[E_0[Y|A=1,W]] = 0.5$$

We varied the treatment mechanism in order to gauge the performance of TMLE and SP-TMLE with different confounding structures and with efficient influence functions that are more or less difficult to approximate as the linear span of different kinds of functions. Below are the different treatment mechanisms we used for $\text{logit}\, g_0(A|W)$:
\begin{align*}
& 2W_2 + W1
& \text{(linear)}
\\ 
& 3*\text{cos}\left(2\pi\sqrt{W_1^2 + W_2^2}\right)
& \text{(sinusoidal)}
\\
& \mathbbm{1}\big((W_1>0 \cap W_2>0) \cup (W_1<0 \cap W_2<0)\big)
& \text{(step)}
\end{align*} 

\subsection{Estimators}

We compared four estimators of the treatment-specific mean: a plug-in highly-adaptive lasso (HAL; see below) estimator, a plug-in relaxed HAL estimator, a TMLE estimator using relaxed HAL as the initial fit, and the SP-TMLE estimator using relaxed HAL as the initial fit and preserving its scores. 

HAL is a nonparametric regression method where the fit is produced by running lasso over a set of saturated 0-order spline bases and their tensor products \cite{hal2016, hal2022}. Among regression methods it stands out because it achieves faster-than $n^{-1/4}$ $\mathcal L_2$ convergence rates in a very large non-parametric function class.\footnote{The set of right-continuous functions of bounded sectional variation} Roughly speaking, HAL converges quickly for all ``near additive'' functions \cite{schuler2024highlyadaptiveridge}. This property makes it appropriate for use in TMLE with tabular data (for which near-additivity is typically a reasonable assumption) since TMLE has a rate requirement on the regression estimators in order to make the remainder term $R$ negligible \cite{tmle2018, imci}. For our purposes, the important thing is that HAL produces estimates of the form $\hat Q_n(1,W) = \text{expit}(\Phi_n(W)^\top \hat\beta$). 

However, since it is a regularized fit, the scores solved by HAL are more complicated to derive than what we have been working with. Therefore for simplicity we work with ``relaxed'' HAL: after obtaining the HAL fit \text{expit}($\Phi_n(W)^\top \hat\beta$), we run a simple unpenalized logistic regression using just the basis functions $\Phi_{n,j}$ corresponding to nonzero coefficients in the HAL fit (by a well-known property of the lasso, only $n$ or fewer coefficients will be nonzero). We therefore ``relax'' the lasso constraint on these coefficients and the resulting estimator with coefficients $\tilde \beta$ exactly solves the score equations corresponding with $h_j(\hat P_n) = \Phi_{n,j}(W) (Y - \Phi_n(W)\tilde\beta)$ for $j$  in the set of nonzero coefficients in the HAL fit \cite{vanderlaan2023adaptivedebiasedmachinelearning}.

In the TMLE estimators the treatment mechanism $g_0(A,W)$ was estimated with a 0-order HAL estimator (in the appendix we present results with a GLM estimator to show what happens under treatment model misspecification). The step size $\delta$ was set to $=0.001$ when implementing the targeted update using the universal least-favorable submodel strategy. 

Each estimator was applied to data of increasing sample size. We repeated our experiments 500 times to obtain the sampling distribution of our estimates. We computed bias, variance, and mean-squared error for each estimator. We obtained coverage for both the TMLE and SP-TMLE by computing confidence intervals using the variance of the estimated influence function after targeting \cite{imci, tmle2018}. Coverage is not shown for the nontargeted plugins because there is no simple and computationally efficient way to do inference for these methods (this is a disadvantage relative to TMLE) \cite{vanderlaan2023adaptivedebiasedmachinelearning,vanderlaan2023higherordersplinehighly}.

\subsection{Results}

\begin{figure}[ht]
\includegraphics[width=16cm]{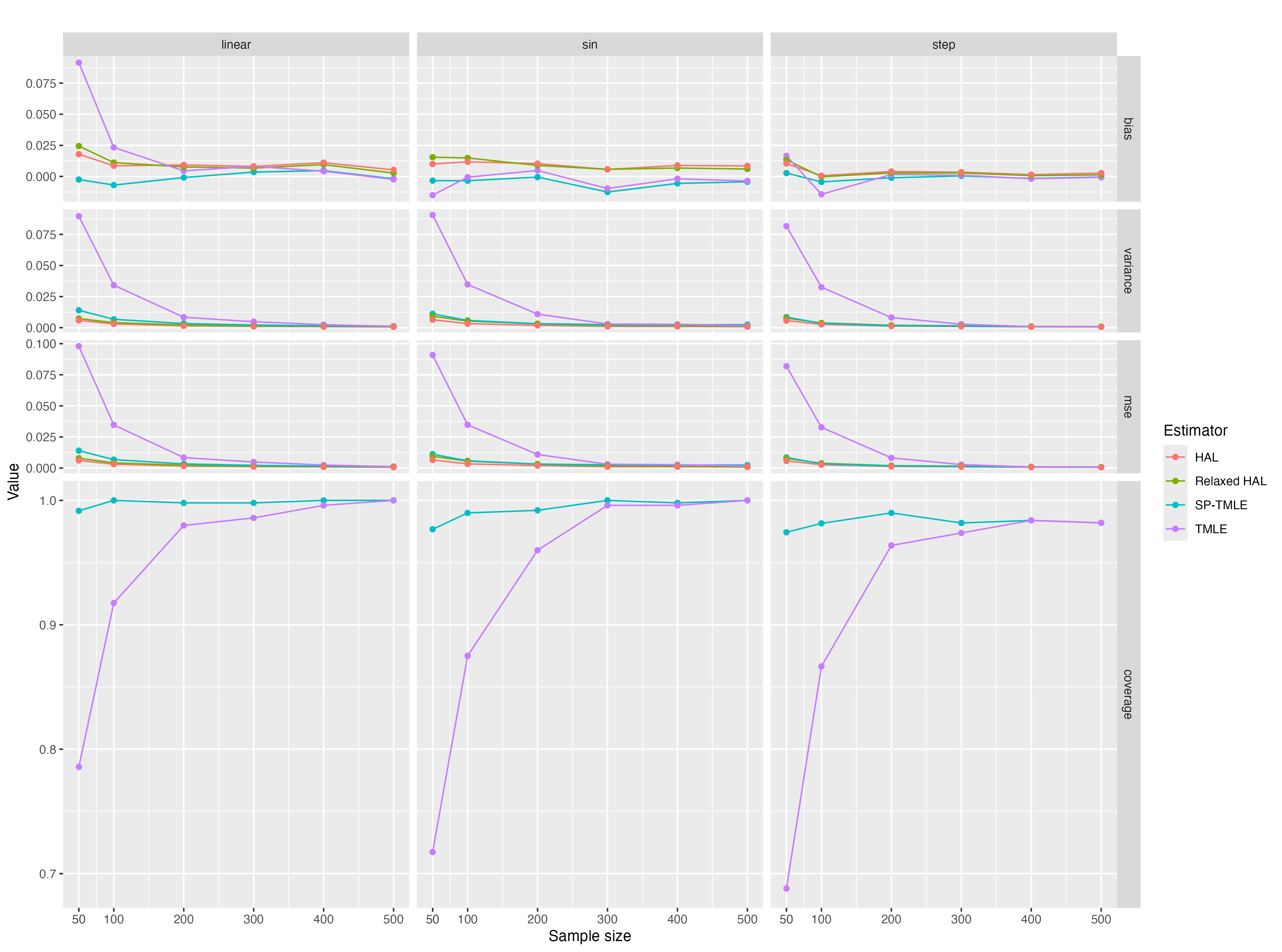}
\centering
\caption[Bias, variance, and MSE of SP-TMLE, TMLE, Relaxed HAL, and HAL estimators]{Bias, variance, and MSE of SP-TMLE, TMLE, relaxed HAL and HAL. Bias was calculated by taking the mean over each 500 estimates of each estimator and subtracting the true parameter value $\Psi(P_0) = 0.5$. Variance was calculated by taking the variance over each 500 estimates of each estimator. 95\% confidence intervals were estimated by taking the empirical variance of the estimated influence function (and dividing by $n$) for each of the 500 estimates, and was used to assess coverage.}
\end{figure}

In small samples, the SP-TMLE estimator beats the standard TMLE estimator. Under each treatment mechanism, we see a substantial reduction in variance relative to TMLE. This is presumably attributable to approximating higher-order efficient influence functions with the span of the scores solved by SP-TMLE and thus reducing the remainder term in display \ref{eq:expansion}. Furthermore, we see that in very small sample sizes the SP-TMLE is better than TMLE at eliminating bias (especially with the linear treatment mechanism).  Again this is attributable to a reduced remainder term on account of solving scores. Both of these improvements contribute to improved coverage for SP-TMLE, as does improved plug-in estimation of the standard error itself (the scores solved by SP-TMLE may begin to span the EIF of the variance of the true influence function).

In this simulation we see that TMLE destroys the good behavior of HAL while SP-TMLE does not. The HAL and relaxed HAL estimator already have good MSE, mainly due to their super-efficient behavior\cite{vanderlaan2023adaptivedebiasedmachinelearning}. However, inference is hard for these estimators as it entails either bootstrapping, which is computationally inefficient, or using the delta method for relaxed HAL \cite{vanderlaan2023adaptivedebiasedmachinelearning}\footnote{We could use the efficient influence function to get inference for plug-in relaxed HAL, acting as if $P_n D^*(\hat{P_n}) \approx0$ is approximately solving it, and it will generally be conservative because HAL acts as a super-efficient estimator.}. In initial testing the latter method was numerically unstable: the LASSO tends to select on the order of $n$ features which often leads to near-singular covariance matrices for the coefficients. However we see that SP-TMLE does provide a bias reduction relative to these estimators, while not sacrificing too large an increase in variance. Therefore SP-TMLE is close to a ``best-of-both-worlds'' option between (possibly super-efficient) plug-in estimators that solve lots of scores and standard TMLE methods. The differences are only apparent in small samples, however.

\section{Discussion}

The SP-TMLE improves the TMLE for small samples. Across the different simulations with varying data-generating processes for the treatment mechanism (linear, sinusoidal, and step function), we saw an improvement in bias for the SP-TMLE estimator in small samples. For our smallest sample size of $n=50$ we see the biggest improvement in bias for the SP-TMLE estimator across each different data-generating process.

In terms of variance, we saw SP-TMLE improving the TMLE. This is due to the SP-TMLE estimator improving the standard error estimator of the variance of the efficient influence function. Across the different simulations for small sample sizes (i.e., $n=50, 100, 200$) the variance of the TMLE estimator was greatest compared to all other estimators and the SP-TMLE estimator. The bias reduction properties of the SP-TMLE estimator and the added benefit of improving the standard error of the variance of the efficient influence function yields better coverage for SP-TMLE relative to TMLE.

A downside of the SP-TMLE is that it couples the TMLE to specific regression estimators (e.g., relaxed HAL in our paper), and one needs to write down the scores explicitly and implement them in the TMLE. Using the relaxed HAL as an initial estimator is convenient and easy to implement, especially because we know exactly the score equations it solves and this lets us implement the SP-TMLE procedure with ease. Our construction generalizes to any kind of ordinary least squares regression with a basis expansion. However, it would be difficult to couple the TMLE procedure with other score-solving estimators unless we had closed-form expressions for the relevant scores. Thus software for SP-TMLEs will be necessarily bespoke unless there is a way to automate or generalize the process for some larger class of initial estimators. 

In higher dimensions we would expect the HAL estimator to be biased relative to the TMLE estimator. In our simulations we do not observe this: HAL is beating the TMLE in most of our cases. We can imagine in higher dimensions a scenario where HAL is losing to TMLE in a more extreme way; however, this is hard to test because because HAL is not computationally scalable in higher dimensions \cite{schuler2023lassoedtreeboosting}. If a higher dimension HAL was computationally feasible, then we would expect for the HAL estimator to be biased relative to the TMLE estimator.

Our work is broadly related to undersmoothing methods, as has already been discussed, and to nonparametric maximum likelihood methods. For example, Kernel Debiased Plug-in Estimation (KDPE) \cite{cho2024kerneldebiasedpluginestimation}, like undersmoothed HAL, solves a rich set of scores and does not rely on \textit{explicitly} solving the efficient influence function. In their work, Cho et al.  \cite{cho2024kerneldebiasedpluginestimation} focus on demonstrating efficient estimation but KDPE may share some of the small-sample benefits of SP-TMLE since the solved scores may be rich enough to span higher-order efficient influence functions.

\newpage
\bibliographystyle{plain}
\bibliography{main}

\newpage
\appendix

\section{Submodel Scores}
Here we briefly verify our claim in the paper that the provided submodel (eq. \ref{eq:sp-update}) appropriately targets both the efficient influence function and the requried scores.

Using equation 4, let us denote \(\text{logit}\, \hat Q_n(\epsilon)\) as \(z(\epsilon)\). Then:
\[
z(\epsilon) = \text{logit}\, \hat Q_n + \epsilon_0 \frac{A}{\hat g_n(W)} + \sum_j^{\tilde d} \epsilon_j \Phi_{n,j}(W).
\]
By solving for \(\frac{d}{d\epsilon_k}\log(\hat Q_n(\epsilon))\) for \(k \in {0,1,\dots,\tilde{d}}\) we can show that that the $(\tilde d + 1)$-dimensional parametric submodel for $\hat Q_n$ in equation 4 spans each of these scores as well as the efficient influence function. First, recall that inverse of the logit function is the the sigmoid function

\[
\hat Q_n(\epsilon) = \text{logit}^{-1}(z(\epsilon)) = \frac{1}{1 + e^{-z(\epsilon)}}.
\]
Now, \(\log\hat Q_n(\epsilon)\) is:
\[
\log\hat Q_n(\epsilon) = \log\left(\frac{1}{1 + e^{-z(\epsilon)}}\right) = -\log(1 + e^{-z(\epsilon)}).
\]
Computing \(\frac{d}{d\epsilon_k} \log\hat Q_n(\epsilon)\) for any parameter \(\epsilon_k\), where \(k = 0\) or \(k \in \{1, \dots, \tilde d\}\):

\begin{enumerate}
    \item First, differentiate \(-\log(1 + e^{-z(\epsilon)})\) with respect to \(z(\epsilon)\):
\[
\frac{d}{dz} \log(\hat Q_n(\epsilon)) = \frac{e^{-z(\epsilon)}}{1 + e^{-z(\epsilon)}} = 1 - \hat Q_n(\epsilon)
\]

    \item Next, apply the chain rule:
    \[
    \frac{d}{d\epsilon_k} \log(\hat Q_n(\epsilon)) = \frac{d}{dz} \log(\hat Q_n(\epsilon)) \cdot \frac{d}{d\epsilon_k} z(\epsilon)
    \]

    \item Compute \(\frac{d}{d\epsilon_k} z(\epsilon)\):
    \begin{itemize}
        \item For \(\epsilon_0\):
        \[
        \frac{\partial z(\epsilon)}{\partial \epsilon_0} = \frac{A}{\hat g_n(W)}
        \]
        \item For \(\epsilon_j\):
        \[
        \frac{\partial z(\epsilon)}{\partial \epsilon_j} = \Phi_{n,j}(W)
        \]
    \end{itemize}

    \item Combine the results:
    \begin{itemize}
        \item For \(\epsilon_0\):
        \[
        \frac{\partial}{\partial \epsilon_0} \log(\hat Q_n(\epsilon)) = (1 - \hat Q_n(\epsilon)) \cdot \frac{A}{\hat g_n(W)}
        \]
        \item For \(\epsilon_j\):
        \[
        \frac{\partial}{\partial \epsilon_j} \log(\hat Q_n(\epsilon)) = (1 - \hat Q_n(\epsilon)) \cdot \Phi_{n,j}(W), \quad j \in \{1, \dots, \tilde d\}
        \]
    \end{itemize}
\end{enumerate}

In the end we have:

\[
\frac{\partial}{\partial \boldsymbol{\epsilon}} \log(\hat Q_n(\epsilon)) = 
\begin{pmatrix}
\frac{\partial}{\partial \epsilon_0} \log(\hat Q_n(\epsilon)) \\
\frac{\partial}{\partial \epsilon_1} \log(\hat Q_n(\epsilon)) \\
\vdots \\
\frac{\partial}{\partial \epsilon_{\tilde{d}}} \log(\hat Q_n(\epsilon))
\end{pmatrix}
= 
\begin{pmatrix}
(1 - \hat Q_n(\epsilon)) \cdot \frac{A}{\hat g_n(W)} \\
(1 - \hat Q_n(\epsilon)) \cdot \Phi_{n,1}(W) \\
\vdots \\
(1 - \hat Q_n(\epsilon)) \cdot \Phi_{n,\tilde{d}}(W)
\end{pmatrix}
\] and we can easily show that this vector spans all of the scores, because some linear combination of them will give us each of the scores. For example, multiplying the k'th element by one and all other k-1 elements by zero.

\end{document}